\documentclass[aip,reprint,amsmath,amssymb]{revtex4-1}

\usepackage[colorlinks=true, urlcolor=blue, linkcolor=blue, citecolor=blue]{hyperref}

\usepackage{graphicx}
\usepackage{dcolumn}
\usepackage{bm}

\usepackage[utf8]{inputenc}
\usepackage[T1]{fontenc}
\usepackage{mathptmx}
\usepackage{siunitx}

\begin{document}

\title{Terahertz photoresistivity of a high-mobility  3D topological insulator based on a strained HgTe film}

\author{M.\,L.\,Savchenko}\thanks{Authors to whom correspondence should be addressed: mlsavchenko@isp.nsc.ru, sergey.ganichev@physik.uni-regensburg.de}
\affiliation{Rzhanov Institute of Semiconductor Physics, 630090 Novosibirsk, Russia}
\affiliation{Novosibirsk State University, 630090 Novosibirsk, Russia}
\author{M.\,Otteneder}
\affiliation{Terahertz Center, University of Regensburg, 93040 Regensburg, Germany}
\author{I.\,A.\,Dmitriev}
\affiliation{Terahertz Center, University of Regensburg, 93040 Regensburg, Germany}
\affiliation{Ioffe Institute, 194021 St. Petersburg, Russia}
\author{N.\,N.\,Mikhailov}
\author{Z.\,D.\,Kvon}
\affiliation{Rzhanov Institute of Semiconductor Physics, 630090 Novosibirsk, Russia}
\affiliation{Novosibirsk State University, 630090 Novosibirsk, Russia}
\author{S.\,D.\,Ganichev}\thanks{Authors to whom correspondence should be addressed: mlsavchenko@isp.nsc.ru, sergey.ganichev@physik.uni-regensburg.de}
\affiliation{Terahertz Center, University of Regensburg, 93040 Regensburg, Germany}

\date{\today}

\begin{abstract}
We report on a detailed study of the terahertz (THz) photoresistivity in a strained HgTe three-dimensional topological insulator (3D TI) for all Fermi level positions: inside the conduction and valence bands, and in the bulk gap.  In the presence of a magnetic field we detect a resonance corresponding to the cyclotron resonance (CR) in the top surface Dirac fermions (DF) and examine the nontrivial dependence of the surface state cyclotron mass on the Fermi level position. We also detect additional resonant features at moderate electron densities and demonstrate that they are caused by the mixing of surface DF and  bulk electrons.  At high electron densities, we observe THz radiation induced $1/B$-periodic low-field magneto-oscillations coupled to harmonics of the CR and demonstrate that they have a common origin with microwave-induced resistance oscillations (MIRO) previously observed in high mobility GaAs-based heterostructures. This observation attests the superior quality of 2D electron system formed by helical surface states in strained HgTe films.
\end{abstract}

\maketitle 

Three-dimensional TIs based on strained HgTe films have been the subject of an intensive study in the last ten years.  This system is a strong topological insulator with electronic properties mediated by conducting surface helical states with close to linear dispersion with spins locked to the electron's momentum~\cite{Fu2007,Dai2008} and is characterized by a very high mobility of the surface DF, reaching $5\times 10^5\,$cm$^2$/Vs in these systems, and low bulk conductivity.  The properties of the surface states have been comprehensively studied using magneto-transport, phase-sensitive SQUID  and capacitance spectroscopy~\cite{Brune2011,Kozlov2014,Brune2014,Maier2015,Kozlov2016,Maier2017,Thomas2017,Ziegler2018,Noel2018,Ziegler2020}. These experiments resulted in the observation of the quantum Hall effect and probing of quantum capacitance in a 3D topological insulator, demonstrated a non-trivial Berry phase of Shubnikov -- de Haas oscillations in transport and capacitance responses, provided an access to a detailed study of the surface states transport properties, and demonstrated highly efficient spin-to-charge current conversion. Presence of the topologically protected conducting surface states in strained HgTe 3D TIs also gives rise to a number of phenomena driven by THz electric fields.  Observation of universal Faraday and Kerr effects~\cite{Shuvaev2012,Dziom2017a} predicted in~Ref.~\onlinecite{Tse2010}; THz quantum Hall effect~\cite{Shuvaev2013}  and photogalvanic currents~\cite{Dantscher2015,Candussio2019} excited in the surface states; study of surface states dynamic applying time domain spectroscopy~\cite{Hancock2011}   and cyclotron resonance spectroscopy~\cite{Shuvaev2013,Dantscher2015,Candussio2019,Gospodaric2019a,Gospodaric2020}, where values of the effective mass of DF from top and bottom surfaces were determined, are only some examples of the achievements in this field. 

While THz radiation induced optical and photocurrent phenomena have been widely investigated there has been no work so far aimed at the study of the photoconductive (photoresistive) response of the surface states. 
Such measurements in 3D TI, however, would not only yield information on tiny details of carrier scattering mechanisms and CR (for HgTe 2D systems see Ref.~\onlinecite{Otteneder2018}) but also may result in the observation of such fascinating phenomena as MIRO previously detected in 2D systems with parabolic dispersion~\cite{Zudov2001,Dmitriev2012} and, most recently, in DF in graphene~\cite{Monch2020}.

In this paper we report on the investigation of the THz photoresistance of HgTe-based 3D TIs for all Fermi level, $E_\text{F}$,  positions: inside the conduction and valence bands, and in the bulk energy gap. Studying the magnetic field dependences of the photoresponse we observed pronounced CR and, at high electron densities, THz radiation induced MIRO-like oscillations coupled to CR. Furthermore, for the intermediate electron densities we detected an additional set of oscillations which behave similarly to magneto-intersubband oscillations (MISO) detected in coupled double quantum wells (QWs)~\cite{Bykov2008a, Wiedmann2008}.

\begin{figure}
	\includegraphics[width=1\columnwidth]{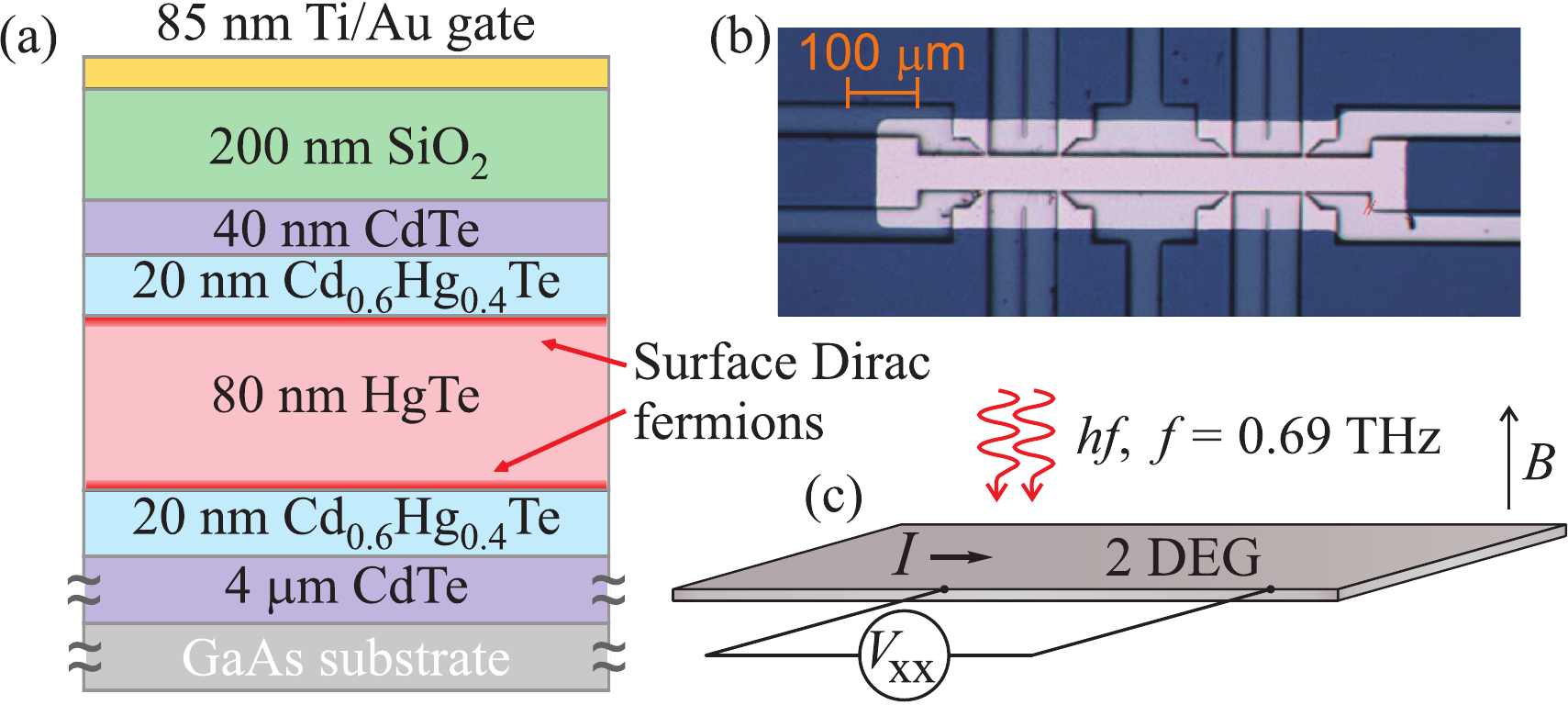}
	\caption{
		(a)~Schematic cross section of the structure under study.  Bright red lines represent surface DF on the top and bottom surfaces of the HgTe film.  (b)~Optical micrograph of the device; faint lilac area corresponds to the gated region. 	
		(c)~Schematic experimental setup.
	} \label{Fig1}
\end{figure}

Experimental samples are field effect transistor-like Hall-bar structures with semi-transparent Ti/Au gates fabricated on the basis of  strained 80-nm HgTe films that have been grown by molecular beam epitaxy on a GaAs (013) substrate~\cite{Kozlov2014}~(Fig.~\ref{Fig1}). The Hall-bar channel width is 50\,$\si{\micro\metre}$ and distances between potential probes are 100 and 250\,$\si{\micro\metre}$.  The samples are placed into an optical cryostat.  We apply a molecular far-infrared laser as a source of THz radiation with frequency $f = 0.69\,$THz (wavelength $\lambda = 432\,\si{\micro\metre}$)~\cite{Kvon2008a,Olbrich2013a,Olbrich2016}. The incident power $P \approx 20\,$mW is modulated at about 160\,Hz by an optical chopper.  Photoresistance is measured by means of a double modulation technique~\cite{Kozlov2011} with a low modulation frequency of 6\,Hz and a high one corresponding to the chopper frequency. The temperature range of the experiment is $(2 - 20)\,$K. 

\begin{figure}
	\includegraphics[width=0.9\columnwidth]{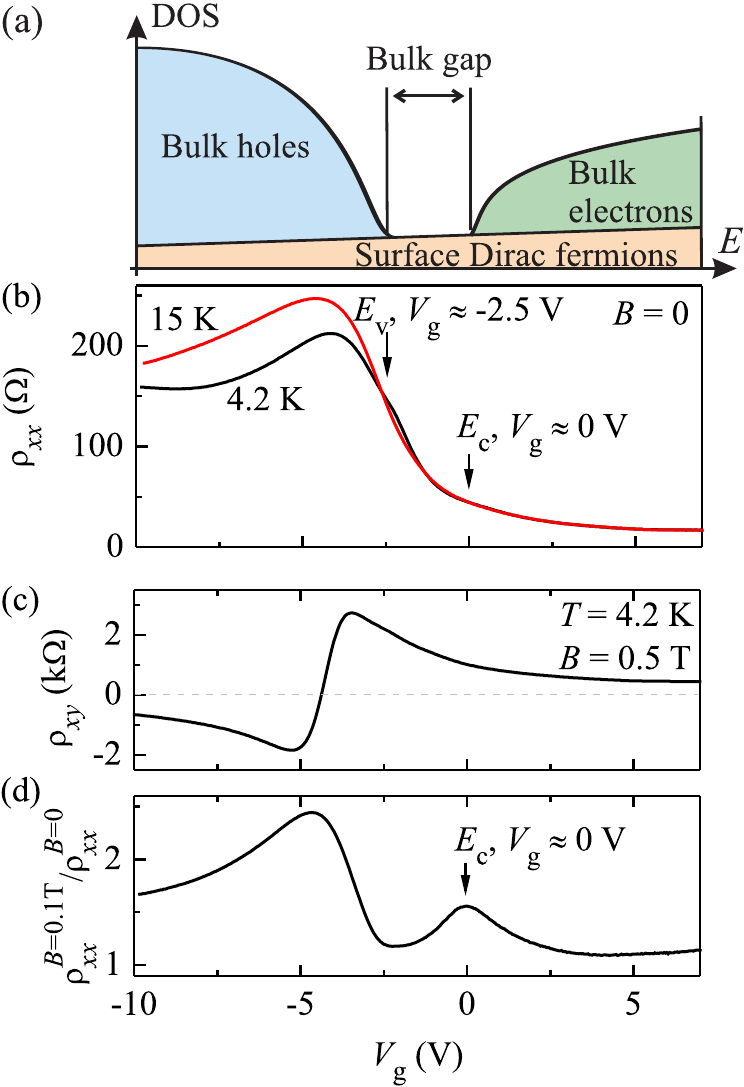}
	\caption{
		(a)~Schematic energy dependence of the density of states (DOS) of the system under study. 		
		(b)~Gate voltage dependence of dissipative resistivity $\rho_{xx}(V_\text{g})$ measured for $B=0$ at $T = 4.2\,$K (black) and $T = 15\,$K (red). 		
		(c)~Gate voltage dependence of the Hall resistivity  $\rho_{xy}(V_\text{g})$ measured at $B = 0.5\,$T and $T = 4.2\,$K.
		(d)~Gate voltage dependence of $\rho_{xx}$ measured at $B = 0.1\,$T and normalized by its value at   zero magnetic field. The plot is used to determine the position of the conduction band bottom.
	} \label{Fig2}
\end{figure}

All studied samples have been characterized by magnetotransport measurements using a standard low-frequency lock-in technique  in a perpendicular magnetic field \textit{B} up to 7\,T and current \textit{I} in the range of $(10 - 100)\,$nA. Typical gate voltage dependences of dissipative, $\rho_{xx}(V_\text{g})$, and Hall, $\rho_{xy}(V_\text{g})$, resistivities are shown in Fig.~\ref{Fig2}\,(b) and (c). 
The values of gate voltage corresponding to the conduction band bottom ($E_\text{c}$) and the valence band top ($E_\text{v}$), see marked arrows in Fig.~\ref{Fig2}, were determined following the methods developed in Refs.~\onlinecite{Kozlov2014} and~\onlinecite{Ziegler2018}. Specifically, the electron-hole scattering, activated by temperature, results in a sharp rise of the zero-field resistivity when  $E_\text{F}$ enters the valence band, see Fig.~\ref{Fig2}\,(b). The low-field magnetoresistance, see Fig.~\ref{Fig2}\,(d), rapidly increases when $E_\text{F}$ enters either the conduction or valence band, reflecting the simultaneous presence of two different types of carriers.

\begin{figure}
	\includegraphics[width=0.9\columnwidth]{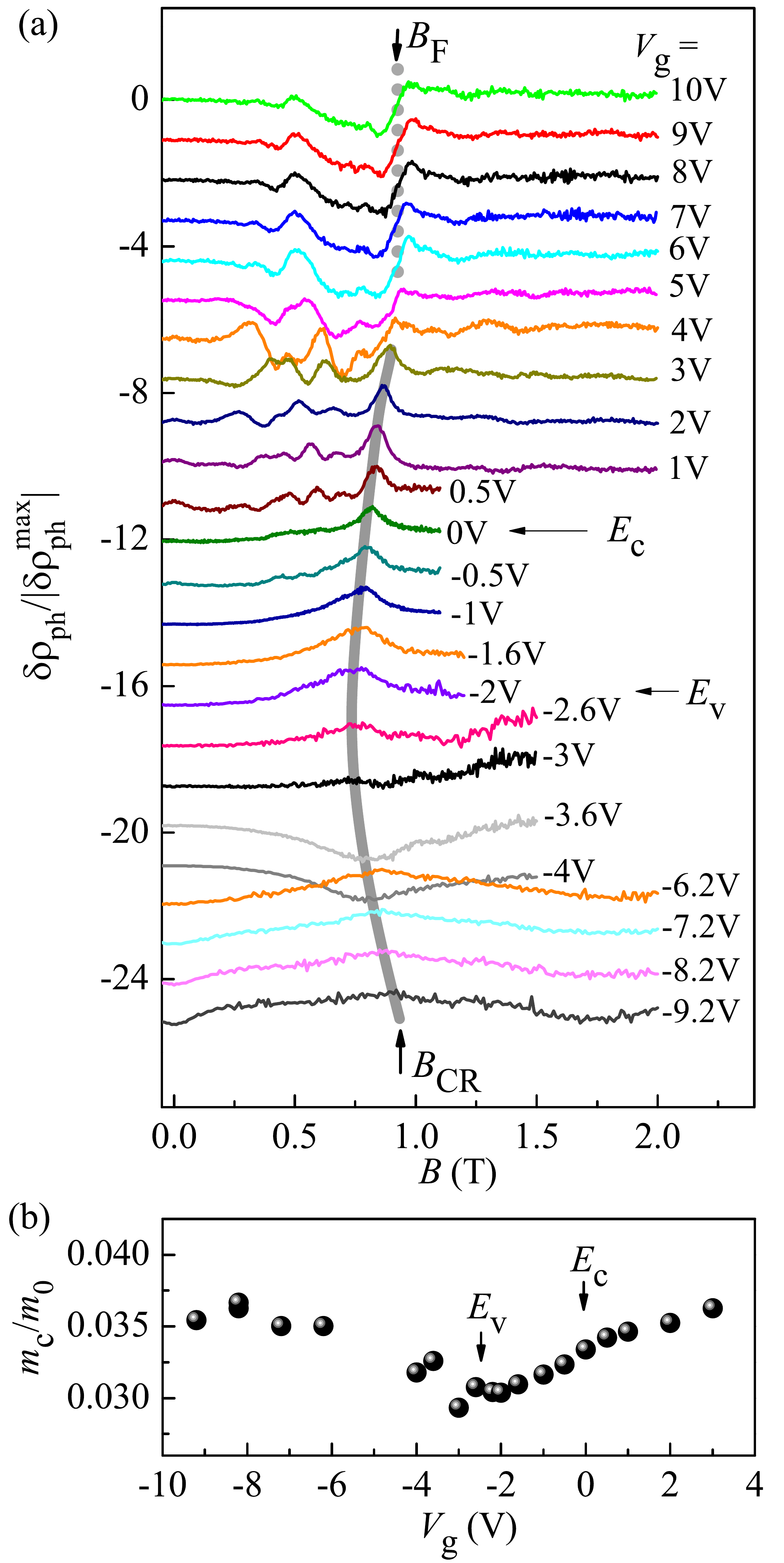}
	\caption{
		(a)~Magnetic field dependences of the normalized photoresistivity $\delta \rho_\text{ph}/|\delta \rho^\text{max}_\text{ph}|(B)$ measured at different gate voltages, corresponding to the position of the Fermi level in the bulk valence band ($V_\text{g} < -2.5\,$V), inside the bulk gap ($-2.5\,\text{V} < V_\text{g} < 0\,$V), and in the bulk conduction band ($V_\text{g} > 0\,$V). 	The curves are vertically shifted by -1.1 for clarity. 	Thick solid and dotted gray lines schematically correspond to the CR positions $B_\text{CR}$ and the fundamental frequency $B_\text{F}$ (see Eq.~(\ref{Eq:3})).
		(b)~Gate voltage dependence of the obtained cyclotron mass.
	} \label{Fig3}
\end{figure}

Fig.~\ref{Fig3}\,(a) demonstrates the main result of our work -- magnetic field dependences of the photoresistivity  $\delta \rho_\text{ph}(B)$ normalized to the maximum of its absolute value $|\delta \rho^\text{max}_\text{ph}|$.  The dependences are  measured at  $\lambda = 432\,\si{\micro\metre}$ for three ranges of $V_\text{g}$ corresponding to the Fermi level position in:  i)  the valence band ($V_\text{g} < -2.5\,$V), ii) in the gap ($-2.5\,\text{V} < V_\text{g} < 0\,$V), and  iii) in the conduction band ($V_\text{g} > 0\,\text{V}$). All curves are measured at 20\,K.  This temperature is high enough to suppress the contribution of the Shubnikov -- de Haas oscillations.  One can clearly see that $\rho_\text{ph}/|\rho^\text{max}_\text{ph}(B)|$ dependences have resonant shapes with the maximum position lying in the magnetic field range $B_\text{CR} = (0.7 - 0.9)\,$T depending on the applied gate voltage, i.e., Fermi level position. Using the value of $B_\text{CR}$ one can determine the cyclotron effective mass $m_\text{c} = e B_\text{CR} /(2 \pi f) $, where $e>0$ is the elementary charge. In Fig.~\ref{Fig3}\,(b) we show the gate voltage dependence of $m_\text{c}$.  One can see that this dependence has a nonmonotonic behavior with a minimum value of $m_\text{c} = 0.03\,m_0$ near the valence band top; $m_0$ is the free electron mass.  The cyclotron mass approaches its maximum value $m_\text{c} = 0.04\,m_0$ at the highest gate voltages, corresponding to a DF density of about $7\times10^{11}\,$cm$^{-2}$.  In fact, such nonmonotonic behavior of the cyclotron mass and its values are in line to what was measured~\cite{Shuvaev2012, Dantscher2015, Gospodaric2019a}, and calculated~\cite{Dantscher2015, Gospodaric2019a} for surface DF in HgTe. 

Measurements of the temperature dependence of the dark resistance at different gate voltages demonstrate that the change of sign of the CR photoresistivity around $V_\text{g} \approx -3$~V in Fig.~\ref{Fig3} is accompanied by the sign change in the correspondent temperature variations of the dark resistivity (see supplementary material). These observations provide a strong support to the conventional heating mechanism of the observed CR photoresistivity.
The photoresistivity in this case can be expressed as
\begin{equation}\label{Eq:1}
\delta \rho_\text{ph} = \alpha A \partial\rho_{xx}(V_\text{g})/\partial T  \,,
\end{equation}
where $\alpha>0$ is a positive coefficient relating the incident power to the increase of the temperature of DF, and $A$ is the absorption coefficient.

\begin{figure}
	\includegraphics[width=1\columnwidth]{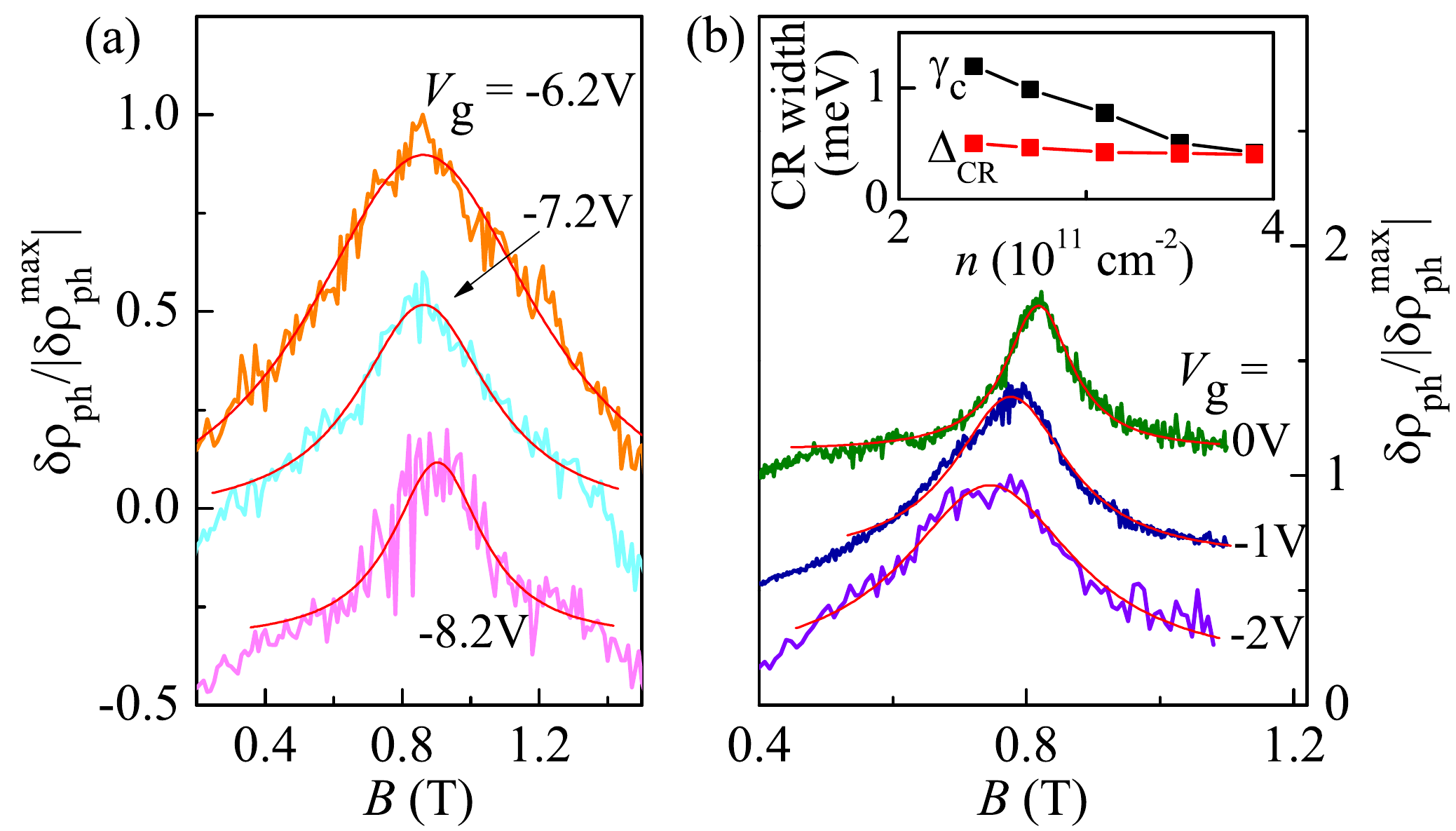}
	\caption{
		Examples of the Lorentzian fitting of the normalized photoresistivity measured when the Fermi level is in the bulk valence band (a) and inside the bulk gap (b). Red curves are Lorentzian fits of CR. \textit{Inset} -- Density dependence of the experimental cyclotron width $\gamma_\text{c}$ (black) and the theoretical cyclotron width for separated Landau levels $\Delta_\text{CR}$ (red) for $E_\text{F}$ positions inside the bulk energy gap.
	} \label{Fig4}
\end{figure}

Now we analyze the shape of the CR photoresistivity.  We begin from a gate voltage range corresponding to the Fermi level positions in the valence band (see Fig.~\ref{Fig4}\,(a)).  In this range the photoresistivity is quite satisfactory fitted by a Lorentzian curve. It is interesting to compare the Lorentzian width $\gamma_\text{c}$ with the theoretical CR width $\Delta_\text{CR}$ for separated Landau levels using an expression for the latter in the case of a short range potential~\cite{Ando1982, Dmitriev2012}
\begin{equation}\label{Eq:2}
\Delta_\text{CR}^2 = 2 \hbar^2 \omega_\text{c} / \pi \tau,
\end{equation}
where $\omega_\text{c}$ is the cyclotron frequency and $\tau$ is the relaxation time. In the valence band the DF mobility is about $2\times10^5\,$cm$^2$/Vs which corresponds to $\Delta_\text{CR} \approx 0.6\,$meV. This value is two to three times larger compared to experimental resonance width $\gamma_\text{c}$ values.  A possible origin of the indicated discrepancy is the inelastic scattering of surface DF by bulk holes which is very significant in the temperature range we used in our experiments~\cite{Kozlov2014}.   

Next we consider the CR photoresistivity shape when the Fermi level enters into the gap (Fig.~\ref{Fig4}\,(b)). The fitting of the photoresistivity by a Lorentzian function also demonstrates quite good agreement.  It gives a resonance width of about 0.3\,T when the Fermi level lies near the top of the  valence band, and it decreases when the Fermi level moves to the bottom of the conduction band.  More careful analysis of the $\gamma_\text{c}$ behavior shows an interesting feature (inset to Fig.~\ref{Fig4}\,(b)):  as the Fermi level moves through the gap from the valence band top to the conduction band bottom a significant CR peak narrowing occurs, while $\Delta_\text{CR}$ has no change. 

\begin{figure}
	\includegraphics[width=0.9\columnwidth]{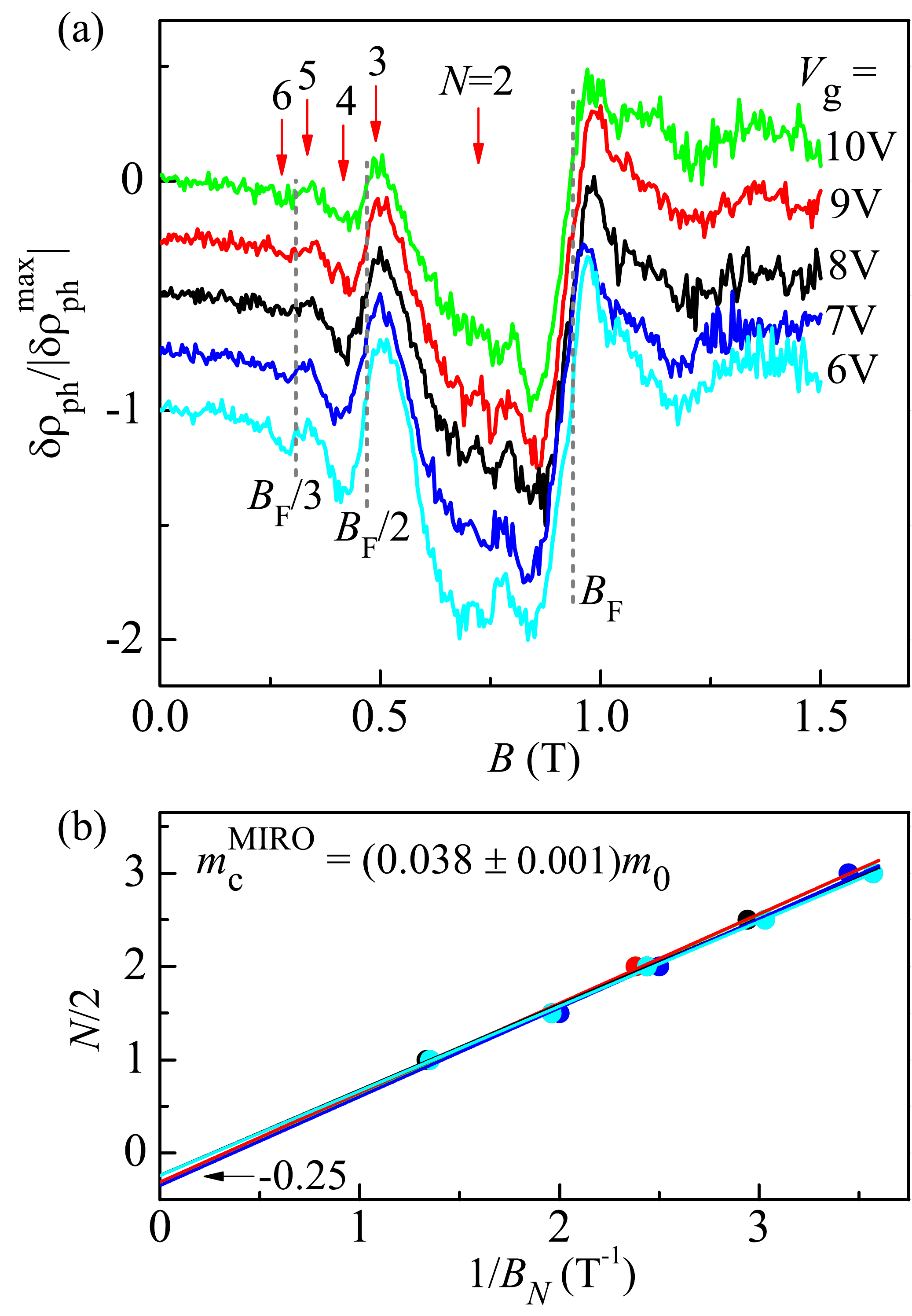}
	\caption{
		(a)~Magnetic field dependences of the normalized photoresistivity $\delta \rho_\text{ph}/|\delta \rho^\text{max}_\text{ph}|$ measured at $V_\text{g} = 6, 7, 8, 9,$ and 10\,V as also shown in Fig.~\ref{Fig3}\,(a). Traces are vertically shifted by -0.25 for clarity.
		(b)~The reciprocal magnetic field dependence of the photoresistivity extrema that are indicated by arrows in panel~(a) and numbered as  $N = 2, 3, \dots$\,.  The symbols and approximation lines have the same color code as the corresponding photoresistivity curves.
	} \label{Fig5}
\end{figure}

As the Fermi level leaves the gap and enters the conduction band (see a $\delta \rho_\text{ph}/|\delta \rho^\text{max}_\text{ph}|(B)$ dependence in Fig.~\ref{Fig3}\,(a) at $V_\text{g} = 0.5\,$V) photoresistivity oscillations at magnetic fields below $B_\text{CR}$ emerge.  Note that these oscillations are absent when the Fermi level intersects only the surface states.  Since the top surface DF have higher density and mobility in the studied systems~\cite{Kozlov2014, Kozlov2016}, the oscillations are presumably generated by the mixing of a top surface DF band and the bulk conduction band, which, more accurately, contains a set of the size-quantized subbands~\cite{Brune2011,Dantscher2015}. So the observed oscillations are similar to magnetointersubband oscillations (MISO) in coupled double QWs studied in~Ref.~\onlinecite{Bykov2008a, Wiedmann2008}.  But in our case we have two significantly different sets of interacted bands: the spin-polarized surface DF band and the size-quantized subbands of the 80\,nm HgTe film. Moreover, in our experiment we have an important advantage: due to the field effect transistor structure we are able to change the position of the Fermi level and correspondingly the subband densities. The absence of MISO at negative voltages is attributed to the much lower mobility and dense spectrum of the bulk holes~\cite{Dantscher2015}.

As the Fermi level moves further inside the conduction band, these oscillations are superimposed with THz induced MIRO-like oscillations that have nearly the same structure for several gate voltages ranging from 6 to 10\,V, see Fig.~\ref{Fig5}\,(a). 
The emergence of MIRO in the photoresistivity at high electron densities is consistent with results of previous studies in different materials~\cite{Dmitriev2012, Tabrea2020, Monch2020}. 
Arrows in Fig.~\ref{Fig5}\,(a) show the positions of subsequent extrema of MIRO, numbered as  $N = 2, 3, \dots$\,. It is well known that for MIRO~\cite{Dmitriev2012}  
\begin{equation}\label{Eq:3}
\delta \rho_\text{ph} \varpropto - \sin(2 \pi B_\text{F}/B), 
\end{equation}
where $ B_\text{F} = 2 \pi f m_\text{c}^\text{MIRO}/e$. Thereby, the slope of the $N/2$ v.s. $B^{-1}$ dependence, see  Fig.~\ref{Fig5}\,(b), is equal to $B_\text{F}$ that allows us to determine the corresponding effective mass $m_\text{c}^\text{MIRO} = (0.038 \pm 0.001)m_0$. This value is close to the cyclotron mass values at high positive $V_\text{g}$, where it is possible to mark out CR ($V_\text{g} = 2$ and 3\,V). We note that approximation lines $N/2 (B^{-1})$ start near $N/2 = -0.25$ as it is established for MIRO~\cite{Dmitriev2012}. A detailed study of the observed transformation of the photoresistance from one CR peak to MIRO-like oscillations through a rich picture of interband interaction induced oscillations will be reported later.

To conclude, we have observed and studied  the THz 
photoresistivity of 80-nm-thick strained HgTe 3D TI.  The photoresistivity was studied at all Fermi level positions: inside the conduction and valence bands and in the bulk gap.  For the Fermi level lying in the valence band or the gap, we observed a single resonance of the photoresistivity, which is caused by the cyclotron resonance of DF in the top surface. For higher positions of the Fermi level, i.e, for $E_\text{F}$ lying in the conduction band, the CR-photoresistivity becomes superimposed fist with magnetointersubband oscillations, and, at further increase of $E_\text{F}$, with MIRO oscillations. The observation of MIRO provides an important evidence of high quality of 2D electron system formed by helical surface states in strained HgTe films.

\section*{Supplementary material} 
See the Supplementary Material bellow.  Content: magnetic field dependencies of the resistivity  measured at different gate voltages and temperatures.

\begin{acknowledgments}

We are grateful to Dima Kozlov for discussions.	
Novosibirsk team acknowledges the financial support by the Russian Science Foundation (Grant No. 16-12-10041-P). Regensburg team gratefully acknowledges the support of the Deutsche Forschungsgemeinschaft (DFG) - Project-ID 314695032 - SFB 1277, and the Volkswagen Stiftung Program (97738). 	S.D.G. also thanks	 the IRAP program of the Foundation for Polish Science (grant MAB/2018/9, CENTERA) for the support.
	
\end{acknowledgments}

\section*{Data availability}
The data that support the findings of this study are available from the corresponding authors upon reasonable request.

\bibliographystyle{aipnum}
\bibliography{library}

\end{document}


\title{Supplementary material to:\\ 
	``Terahertz photoresistivity of a high-mobility  3D topological insulator based on a strained HgTe film''}

\author{M.\,L.\,Savchenko}\thanks{Authors to whom correspondence should be addressed: mlsavchenko@isp.nsc.ru, sergey.ganichev@physik.uni-regensburg.de}
\affiliation{Rzhanov Institute of Semiconductor Physics, 630090 Novosibirsk, Russia}
\affiliation{Novosibirsk State University, 630090 Novosibirsk, Russia}
\author{M.\,Otteneder}
\affiliation{Terahertz Center, University of Regensburg, 93040 Regensburg, Germany}
\author{I.\,A.\,Dmitriev}
\affiliation{Terahertz Center, University of Regensburg, 93040 Regensburg, Germany}
\affiliation{Ioffe Institute, 194021 St. Petersburg, Russia}
\author{N.\,N.\,Mikhailov}
\author{Z.\,D.\,Kvon}
\affiliation{Rzhanov Institute of Semiconductor Physics, 630090 Novosibirsk, Russia}
\affiliation{Novosibirsk State University, 630090 Novosibirsk, Russia}
\author{S.\,D.\,Ganichev}\thanks{Authors to whom correspondence should be addressed: mlsavchenko@isp.nsc.ru, sergey.ganichev@physik.uni-regensburg.de}
\affiliation{Terahertz Center, University of Regensburg, 93040 Regensburg, Germany}


\maketitle 

\setcounter{figure}{0}
\renewcommand{\thefigure} {S\arabic{figure}}

\begin{figure}
	\includegraphics[width=1\columnwidth]{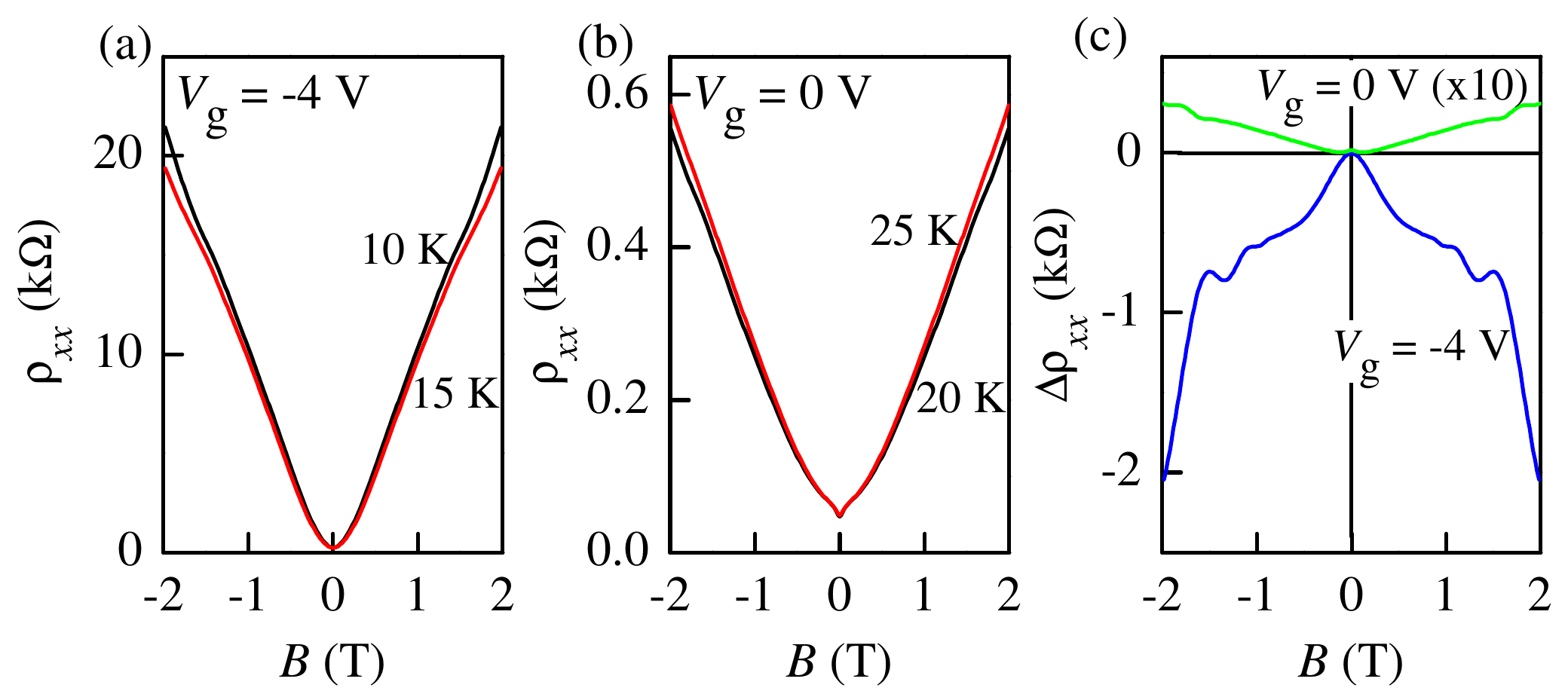}
	\caption{
		(a)~Magnetic field dependence of the resistivity measured at $T = 10\,$K (black) and at $T = 15\,$K (red) for $V_\text{g} = -4\,$V. 
		(b)~Magnetic field dependence of the resistivity measured at $T = 20\,$K (black) and at $T = 25\,$K (red) for $V_\text{g} = 0\,$V. 
		(c)~Magnetic field dependence of the differential resistivity $\Delta \rho_{xx} = \rho_{xx}^\text{higher T} - \rho_{xx}^\text{lower T}$ for $V_\text{g} = -4\,$V (blue) and $V_\text{g} = 0\,$V (green, multiplied by ten).
	} \label{FigS1}
\end{figure}

The sign of the observed cyclotron photoresistance $\delta\rho_\text{ph}$ (see Fig.~3\,(a) of the main text) is positive for all gate voltages except $V_\text{g} = -3.6$ and -4\,V.
In the manuscript we propose that carrier heating is responsible for the photoresistivity, and following Eq.~(1) of the manuscript the sign of the photoresistivity is determined by the sign of the temperature derivative of resistance $\partial \rho_{xx}/\partial T$. 
In~Fig.~\ref{FigS1} we show the examples of the magnetic field dependence of the resistivity $\rho_{xx}(B)$ measured at two temperatures at $V_\text{g}= - 4\,$V~(a) and at $V_\text{g} = 0\,$V~(b).
We observe positive classic magnetoresistance at both gate voltages arising from the coexistence of several types of carriers in the studied system at all $V_\text{g}$~\cite{Kozlov2014}.
But at $V_\text{g} = -4\,$V the Fermi level is near the charge neutrality point, where there is an equal number of the Dirac surface electrons and bulk holes, while at $V_\text{g} = 0\,$V there are only electrons present in the system.
This results in the huge difference in the magnitude of the magnetoresistance $\rho_{xx}(B)/\rho_{xx}(B=0)$:
it reaches about 10\,000\% for panel (a) and about 1\,200\% panel (b) at $B = 2\,$T.
Because of the different states in the system the temperature dependences of density and mobility of carriers, which govern the $\rho_{xx}(B)$ dependence, can also be different~\cite{Kozlov2014}.
This leads to opposite signs of $\partial \rho_{xx}/\partial T$ for these gate voltages which we emphasize in~Fig.~\ref{FigS1}\,(c) where we show the difference $\Delta \rho_{xx} = \rho_{xx}^\text{higher T} - \rho_{xx}^\text{lower T}$ for $V_\text{g} = -4\,$V (blue, negative sign) and $V_\text{g} = 0\,$V (green, multiplied by ten, positive sign). These results fully support the heating mechanism of the observed CR photoresistivity.

\bibliographystyle{aipnum}
\bibliography{library}